\documentclass[12pt]{article}
\usepackage{a4}
\usepackage{amsfonts,amssymb,amsmath,amsthm}
\usepackage{graphicx}
\usepackage{cite}

\begin{document}
\title{Quantum corrections to the noncommutative kink}
\author{R. A. Konoplya\\
\emph{Department of Physics, Kyoto University, Kyoto 606-8501, Japan}\\
and \\
D.~V.~Vassilevich\thanks{On leave from V.~A.~Fock Institute of
Physics, St.~Petersburg University, Russia.
E.mail:\ {\texttt{dmitry(at)dfn.if.usp.br}}}\\
{\it Instituto de F\'isica, Universidade de S\~ao Paulo,}\\ {\it
Caixa Postal 66318 CEP 05315-970, S\~ao Paulo, S.P., Brazil}}

\maketitle

\begin{abstract}
We calculate quantum corrections to the mass of noncommutative 
$\phi^4$ kink
in $(1+1)$ dimensions for intermediate and large values of the
noncommutativity parameter $\theta$. All one-loop divergences are
removed by a mass renormalization (which is different from the one
required in the topologically trivial sector). For large $\theta$ quantum
corrections to the mass grow linearly with $\theta$ signaling
about possible break down of the perturbative expansion.
\end{abstract}

\section{Introduction}\label{sec-intro}
The problem of how to compute quantum corrections to the mass
of $(1+1)$ dimensional solitons was first addressed by Dashen,
Hasslacher and Neveu \cite{Dashen:1974cj} in 1974. This work
was followed by a decade of activities related to quantum
properties of solitons, see \cite{Rajaraman,FK}. The subject
was re-opened in 1997 in the context of supersymmetric solitons
\cite{Rebhan:1997iv,Nastase:1998sy,Graham:1998qq,Shifman:1998zy}
(see also reviews \cite{Rebhan:2004vu,Shifman:2007ce}). About the same
time many methods of calculations appeared, the ones based on the
heat kernel expansion are most important for us
\cite{Bordag:1994jz,Dunne:1999du,Bordag:2002dg,AlonsoIzquierdo:2002eb,
Izquierdo:2006ds}.

Previous studies of solitons in noncommutative (NC) theories 
(cf. \cite{Douglas:2001ba,
Lechtenfeld:2006iz}) avoided the problem of quantum
corrections to their mass, except for the work \cite{Kurkcuoglu:2007rr},
where quantum correction to an NC sine-Gordon soliton were
calculated in a couple of leading orders of the $\theta$-expansion.
This paper also reported certain difficulties in calculating the mass
shift, which are probably not specific to the model considered, but
rather common to all time-space NC theories. 

The present paper is devoted to calculation of quantum corrections
to the mass of NC kink in (1+1) dimensions. We do not use the
$\theta$-expansion. On the contrary, we concentrate on finite and
large values of the noncommutativity parameter. As in \cite{Kurkcuoglu:2007rr}
we define the mass shift (the vacuum energy $E$) as, roughly speaking,
half the sum of eigenfrequencies of fluctuations about the soliton.
Such a definitions is hard to justify rigorously since
in time-space NC theories 
where there is no canonical definition of energy.
Some arguments in favor of treating eigenfrequencies as 
one particle energies can be found in \cite{Strelchenko:2007xh}.
Anyway, $E$ is useful for the discussion of quantum properties of
solitons and renormalization.

This paper is organized as follows. The next section contain some preliminary
material on the NC kink and fluctuations about it mostly taken from
\cite{Vassilevich:2003he}. In sec. \ref{sec-qe} we study spectral
density of the fluctuation and renormalization of the vacuum energy
(which is done by adding a mass counterterm to the classical action).
Section \ref{sec-cal} is devoted to numerical calculations of the
vacuum energy for several values of the NC parameter $\theta=2,3,5,7,10$
(in some natural units). 
The bound state frequencies are evaluated with the help of the WKB
method, while the contribution of the continuous spectrum is calculated by
using an approximating square well potential. We find that quantum
corrections to the vacuum energy grow linearly with $\theta$.
\section{Classical solution and fluctuations}\label{sec-sol}
We take the classical action of NC $\phi^4$ model in $1+1$ dimensions
in the form
\begin{equation}
S=\int d^2x \left[ \frac 12 (\partial_\mu \phi)(\partial^\mu\phi)
 - \frac {\lambda}4 \left( \phi^2_\star -\frac {m^2}\lambda \right)^2 \right],
\label{clac}
\end{equation}
where $\phi^2_\star \equiv \phi\star\phi$. Star
denotes the Moyal product
\begin{equation}
(f\star g) (x)=\left[ \exp \left( \frac i2 \Theta^{\mu\nu} \partial_\mu^x
\partial_\nu^y \right)f(x)g(y) \right]_{y^\mu=x^\mu},\label{Moyal}
\end{equation}
where $\Theta^{\mu\nu}$ is a constant skew-symmetric matrix which 
can be chosen as
$\Theta^{\mu\nu} = 2\theta \epsilon^{\mu\nu}$ with $\epsilon^{01}=1$.
After splitting the coordinates into time and space, $\{ x^\mu\}=\{t,x\}$,
we have the following useful formulae
\begin{equation}
f(x) \star e^{i\omega t}=e^{i\omega t} f(x+\theta \omega),\quad
e^{i\omega t}\star f(x) =e^{i\omega t} f(x-\theta \omega)
\label{shift}
\end{equation}

Obviously, static solutions of the commutative $\phi^4$ remain also
solutions in the NC case. In particular, we have the kink solution
\begin{equation}
\Phi (x)= \frac {m}{\sqrt{\lambda}} \tanh \left( \frac {mx}{\sqrt{2}} \right).
\label{kink}
\end{equation}
Classical energy of this soliton is
\begin{equation}
E_{\rm cl}=\frac {2\sqrt{2}}{3}\, \frac{m^3}\lambda \,.\label{clen}
\end{equation}

Qualitatively the spectrum of fluctuations above the kink solution in NC
$\phi^4$ theory was studied in \cite{Vassilevich:2003he}. Here we repeat
some steps from that paper. Let us split $\phi = \Phi + \varphi$ with 
$\Phi$ being the kink solution (\ref{kink}) and $\varphi$ describing
quantum fluctuations. The linearized equation of motion reads
\begin{equation}
-\partial_t^2 \varphi +\partial_x^2 \varphi + m^2 \varphi 
-\lambda (\Phi \star \Phi \star \varphi + \Phi \star \varphi \star \Phi
+\varphi \star \Phi \star \Phi )=0.\label{lin1}
\end{equation}
Since the kink (\ref{kink}) is static, we can look for the solutions
of (\ref{lin1}) in the form $\varphi = e^{i\omega t}\eta (x)$.
By substituting this ansatz in (\ref{lin1}) and using the relations
(\ref{shift}) we obtain the following equation for $\eta$
\begin{equation}
\omega^2 \eta + \partial_x^2 \eta + m^2\eta - \lambda (
\Phi_+^2 + \Phi_+\Phi_-+\Phi_-^2)\eta =0 \,,\label{lin2}
\end{equation}
where $\Phi_{\pm}(x)=\Phi (x_\pm)$ and $x_\pm = x\pm \theta \omega$. 
It is convenient to choose the mass units such that
\begin{equation}
m=\sqrt{2}.\label{units}
\end{equation}
Then eq.\ (\ref{lin2}) yields
\begin{equation}
(-\partial_x^2 + M^2 + U(x;\omega))\eta = \omega^2 \eta \,,\label{lin3}
\end{equation}
where 
\begin{equation}
U(x;\omega)=2(\tanh^2(x_+)+\tanh(x_+)\tanh (x_-)+\tanh^2 (x_-)-3)\label{Ux}
\end{equation}
and the constant part $M^2=4$ is selected in such a way that 
$U(x;\omega)\to 0$ exponentially fast for $x\to \pm \infty$ and a fixed
$\omega$.

\section{Quantum energy of the fluctuations}\label{sec-qe}
The vacuum energy can be formally defined as a sum of eigenfrequencies
of quantum fluctuations $E=\frac \hbar{2} \sum \omega$. In NC theories
with time-space noncommutativity (in particular, in all NC theories in
$(1+1)$ dimensions) it is hard to define a canonical
Hamiltonian (due to the presence of an infinite number of time
derivatives) and thus to justify this formula for $E$ rigorously. 
Here we follow \cite{Kurkcuoglu:2007rr} and accept this definition
of the vacuum energy (see the comments made in sec.\ \ref{sec-intro}
above).
More precisely, after taking $\hbar =1$ we split the vacuum energy as
\begin{equation}
E=E_B+E_C, \label{evac}
\end{equation}
where 
\begin{equation}
E_B=\frac 12 \sum \omega_B \label{EB}
\end{equation}
is a finite sum over the bound state frequencies. The contribution of the
continuous spectrum has to be regularized. In the zeta function regularization
it reads
\begin{equation}
E_C=\int_M^\infty d\omega\, \omega^{1-2s} \rho (\omega) \,,\label{EC1}
\end{equation}
where a regularization parameter $s$ has been introduced which should
be put zero at the end of the calculations. The function $\rho (\omega)$ is the
spectral density in the continuous spectrum. By using the well known
relation between the spectral density and the phase shift we can
write
\begin{equation}
E_C=\frac 1{2\pi} \int_0^\infty dk\, (k^2+M^2)^{\frac 12 -s} \partial_k
\delta (k), \label{EC2}
\end{equation}
where $\omega = \sqrt{k^2+M^2}$. (For the sake of completeness we rederive
this formula for noncommutative case in Appendix \ref{app-ve}).
Equivalently,
\begin{equation}
E_C=\frac 1{2\pi} \int_M^\infty d\omega\, \omega^{1-2s} \partial_\omega
\delta (\omega) .\label{EC3}
\end{equation}
This quantity is divergent in the limit $s\to 0$. To get rid of the
divergences one has to know the the asymptotic behavior of the
spectral density for large frequencies.

\subsection{Asymptotic behavior of the spectral density}\label{sec-spd}
Since the potential in (\ref{lin3}) depends on the frequency $\omega$
we are dealing with a {\em non-linear spectral problem}. To analyze the
spectral density we use a method developed initially in 
\cite{Fursaev:2000dv,Fursaev:2001fm,Fursaev:2002vi}
and then adapted to NC theories in \cite{Strelchenko:2007xh}.

Let us consider an auxiliary eigenvalue problem
\begin{equation}
L(\sigma)\psi_{\sigma,\omega}=\omega^2\psi_{\sigma,\omega},
\label{aux}
\end{equation}
where
\begin{equation}
L(\sigma)=-\partial_x^2 + M^2 + U(x;\sigma).\label{Lsi}
\end{equation}
The functions $\eta_\omega=\psi_{\omega,\omega}$ solve our initial 
eigenvalue problem (\ref{lin3}), but the spectral densities are different.
Let us denote the density of the auxiliary problem by $\rho (\sigma,\omega)$.
Note, that according to Appendix \ref{app-ve} we are actually working
with the densities from which (an infinite) contribution from free massive
fields in an infinite space has been subtracted. By using this density, one
can calculate spectral functions of the operators $L(\sigma)$ (with
a fixed $\sigma$). In particular, the heat kernel for $L(\sigma)$
reads
\begin{equation}
K(L(\sigma);t)={\rm Tr} (e^{-tL(\sigma)} - e^{-t(-\partial_x^2 +M^2)})
=\int d\omega\ e^{-t\omega^2} \rho (\sigma,\omega). \label{hk}
\end{equation}
Here ${\rm Tr}$ denotes a trace over the space of square integrable functions
on the real line. Again, a subtraction of the ``free'' heat kernel
is necessary. Note, that the integration in (\ref{hk}) must be extended
over the whole spectrum, including the bound states.

In turn, the spectral density can be expressed through the heat kernel
by means of an inverse Laplace transformation.

For any fixed real $\sigma$ the operator $L(\sigma)$ is a Laplace type
operator with a smooth potential. Therefore, the following asymptotic
expansion\footnote{For recent reviews on
the heat kernel expansion the reader can consult \cite{Vassilevich:2003xt}
in the commutative case, and \cite{Vassilevich:2007fq}
on NC spaces.} 
is valid as $t\to +0$
\begin{equation}
K(L(\sigma);t)\simeq \sum_{n=1}^\infty t^{n -1/2} a_{2n} (\sigma).
\label{aymptotex}
\end{equation}
Odd-numbered coefficients vanish since there is no boundary. 
The coefficient $a_0$
does not contribute because of the subtraction of a free heat kernel
in (\ref{hk}). Two leading coefficients read
\begin{eqnarray}
&&a_2(\sigma)=-(4\pi)^{-1/2}\int dx\, U(x;\sigma),\label{a2}\\
&&a_4(\sigma)=(4\pi)^{-1/2}\int dx\,\left[ \frac 12 U(x;\sigma)^2
+M^2 U(x;\sigma)\right].\label{a4}
\end{eqnarray}
In our case, by using (\ref{Ux}) we obtain
\begin{equation}
a_2(\sigma)=\frac 4{\sqrt{\pi}} (\theta \sigma \coth (2\theta\sigma) +1).
\label{a2U}
\end{equation}
For large $\sigma$ we have
\begin{equation}
a_2(\sigma)=\sigma a_{2,1} + a_{2,0} +{\rm e.s.t.}\label{a2si}
\end{equation}
where
\begin{equation}
a_{2,1}=\frac{4\theta}{\sqrt{\pi}},\qquad 
a_{2,0}=\frac{4}{\sqrt{\pi}},
\label{a210}
\end{equation}
and the corrections in eq.\ (\ref{a2si}) are exponentially small.
For the future use we note 
\begin{equation}
a_{2,0}=-\frac \lambda{\sqrt{\pi}} \int dx \, \left( \Phi^2 - 
\frac {m^2}\lambda \right) \,.\label{a2ki}
\end{equation}
Here we used explicit form of the kink solution (\ref{kink})
and restored the $m$-dependence on dimensional grounds.

Higher heat kernel coefficients $a_{2p}$ are integrals of local
polynomials constructed from the potential $U(x;\sigma)$ and its'
derivatives. One can easily prove that for large $\sigma$
\begin{equation}
a_{2p}(\sigma)=\sigma a_{2p,1}+a_{2p,0}+\mathcal{O}(1/\sigma).\label{las}
\end{equation}
Probably, the corrections above are even exponentially small, but we
shall not rely on this. 

The spectral density $\rho (\sigma,\omega)$ taken at coinciding arguments
is {\it not} the density $\rho (\omega)$ of our initial spectral problem
(\ref{lin3}). As demonstrated in \cite{Fursaev:2000dv,Fursaev:2001fm,
Fursaev:2002vi} (see also \cite{Strelchenko:2007xh} for a discussion
in the framework of
NC theories) one has to construct another density $\varrho (\sigma,\omega)$
which is related to a heat-kernel like object
\begin{equation}
\tilde K(\sigma;t)=\left( 1 + \frac 1{2\sigma t} \, \frac {\partial}
{\partial \sigma} \right) K(L(\sigma);t) \label{tilK}
\end{equation}
through the equation
\begin{equation}
\tilde K(\sigma;t)=\int d\omega \, \varrho (\sigma,\omega)\, e^{-t\omega^2}.
\label{tKvr}
\end{equation}
Then
\begin{equation}
\rho (\omega)=\varrho (\omega,\omega).
\label{rr}
\end{equation}

We do not know any differential or pseudo-differential 
operator $\tilde L(\sigma)$ such that $\tilde K(\sigma;t)=
{\rm Tr}(e^{-t\tilde L(\sigma)})$. In any case, such $\tilde L$
cannot be a Laplacian on the real line with a smooth potential.
However, both $\tilde K$ and $\varrho$ are well defined, which
allows us to consider other spectral functions. 

For $t\to +0$
\begin{eqnarray}
&&\tilde K(\sigma;t)\simeq \sum_{n=0}^\infty \tilde a_{2n}(\sigma)
\, t^{-\frac 12 +n},\nonumber\\
&&\tilde a_{2n}(\sigma)=a_{2n}(\sigma)+\frac 1{2\sigma}
a_{2n+2}(\sigma).\label{tila}
\end{eqnarray}
For large $\sigma$ we have
\begin{equation}
\tilde K(\sigma;t)\simeq t^{-1/2} \frac 1{2\sigma} a_{2,1}
+t^{1/2} \left( \sigma a_{2,1} + a_{2,0} +  \frac 1{2\sigma} a_{4,1}
\right) + \dots \label{exptK}
\end{equation}
We are going to use this expansion to evaluate the large $\omega$ behavior
of $\varrho (\sigma,\omega)$, and, after setting $\sigma=\omega$ - the large
$\omega$ behavior of the physical density $\rho (\omega)$. 
As noted in \cite{Fulling97,Fursaev:2002vi}, the problem is that besides from
powers of $\omega$ the high frequency asymptotics of the spectral density 
contain oscillating terms which are not defined by the heat trace 
asymptotics. Strictly speaking, the heat kernel expansion defines the
asymptotic behavior of the so-called Riesz means of the spectral density
rather than that of the spectral density itself. This is not precisely
what we need, but we learn an important lesson: the power-law asymptotics
{\it are} defined by the heat kernel expansion. Oscillating terms are
less important anyway since corresponding contributions to the
vacuum energy are better convergent than that from pure powers.

Let us consider the zeta function corresponding to $\varrho (\sigma,\omega)$
\begin{equation}
\tilde \zeta (\nu)=\int d\omega\, (\omega^2)^{-\nu} \varrho (\sigma,\omega).
\label{tilz}
\end{equation}
One should be careful with possible contribution from $\omega=0$. One should
either treat such state separately, or add a small positive part to the
mass. The details are not essential for us since we are interested in
the behavior at large $\omega$.

There is a well know relation between residues of the zeta function and
the heat kernel coefficients 
\begin{equation}
\tilde a_{2k}={\rm Res}_{\nu = \frac 12 -k} \Gamma (\nu)\tilde \zeta (\nu)=
\Gamma \left( \frac 12 -k \right) {\rm Res}_{\nu = \frac 12 -k}
\tilde \zeta (\nu) .\label{aRes}
\end{equation}
On the other hand, if the spectral density has a contribution behaving
like $c_p \omega^{p}$ at large $\omega$, the zeta function has a pole
term
\begin{equation}
\sim 2\int _{\Omega}^\infty c_p \omega^{p-2\nu} \sim
\frac {c_p}{\nu -(p+1)/2}\,, \label{pole}
\end{equation}
where $\Omega$ is a large number, and the coefficient of $2$ appeared
because we have to take into account degeneracy of the continuous spectrum.

Oscillatory terms do not contribute to the poles. Indeed,
after analytical continuation from large positive $\nu$
the integral $\int_\Omega^\infty d\omega \omega^{-\nu}
\sin (b\omega)$ has no poles on the real line.

Next we compare (\ref{pole}) with (\ref{aRes}) and (\ref{tila}),
(\ref{exptK}) to
obtain the following power law asymptotics of the spectral density
\begin{equation}
\varrho (\sigma,\omega)\simeq \frac 1{2\sqrt{\pi}\sigma} a_{2,1}
-\frac 1{2\sqrt{\pi}} \, \omega^{-2} (a_{2,1}\sigma + a_{2,0}
+a_{4,1}/(2\sigma)) + \dots \label{asvrho}
\end{equation}
For the physical spectral density we have
\begin{equation}
\rho(\omega)=\varrho (\omega,\omega)\simeq -\frac 1{2\sqrt{\pi}} 
\omega^{-2} a_{2,0} + \mathcal{O} (\omega^{-3}) \label{asrho}
\end{equation}
Note, that all terms with $\omega^{-1}$ cancel against each 
other\footnote{The terms with $1/\omega$ in the spectral density lead
to linear divergences in the vacuum energy. In the zeta-function
regularization such divergences are removed automatically, but they
may cause problems in other regularization schemes.}.
This formula does not contain possible oscillating term which
cannot be obtained by these methods.

One should not be afraid of negative spectral densities. We have subtracted
the spectral density of a free massive field. What remains can have both
signs.
\subsection{Renormalization}
Let us now consider renormalization of the vacuum energy. Since only the
contribution from continuous spectrum is divergent, we treat this term only
\begin{equation}
E_C=\mu^{2s}
\int_M^\infty d\omega\, \omega^{1-2s} \rho (\omega) \,,\label{ECmu}
\end{equation}
where we introduced a parameter $\mu$ with the dimension of mass in order
to keep proper dimension of the vacuum energy $E_C$ independently of
the regularization parameter $s$. Next we choose some frequency $\Omega$
and split the integral into two parts. The part from $M$ to $\Omega$
we leave as it is. In the part from $\Omega$ to infinity we add and
subtract the asymptotics $\rho_{\rm as}(\omega)$ of the spectral
density.
\begin{eqnarray}
&&E_C=\mu^{2s}
\int_M^\Omega d\omega\, \omega^{1-2s} \rho (\omega)+
\mu^{2s}\int_\Omega^\infty d\omega\, \omega^{1-2s} 
(\rho (\omega)-\rho_{\rm as}(\omega))\nonumber\\
&&\qquad +
\mu^{2s}
\int_\Omega^\infty d\omega\, \omega^{1-2s} \rho_{\rm as} (\omega) \,.
\label{ECadsub}
\end{eqnarray}
The choice of $\Omega$ is simply a matter of convenience. 
All ultraviolet divergences are contained in the last term. As 
$\rho_{\rm as}$ we take the first term in (\ref{asrho}),
\begin{equation}
\rho_{\rm as}(\omega)=-\frac 1{2\sqrt{\pi}} \,
\omega^{-2} a_{2,0} =-\frac 2\pi \omega^{-2},\label{rhoas}
\end{equation}
where we took into account (\ref{a210}). We assumed that oscillating
terms in the spectral densities (which are beyond our control) do not
contribute to the divergences. This assumption cannot be universally true.
However, in the case we consider in this paper the subtraction 
of (\ref{rhoas}) indeed gives a convergent integral (see below). Therefore,
the assumption we made is correct as well as the renormalization of the
vacuum energy which will be done in a moment. 

The last term in (\ref{ECadsub}) can be easily evaluated,
\begin{equation}
E_{\rm div}=\mu^{2s}
\int_\Omega^\infty d\omega\, \omega^{1-2s} \rho_{\rm as} (\omega)=
-\frac 1{2\sqrt{\pi}} \, a_{2,0} \, \frac 1{2s} \left( \frac \mu\Omega
\right)^{2s}\label{Ediv}
\end{equation}
Near $s=0$ $E_{\rm div}$ behaves as 
\begin{equation}
E_{\rm div} = -\frac 1{2\sqrt{\pi}} \, a_{2,0} \left[ \frac 1{2s}+
\ln \left(\frac \mu\Omega\right ) + \mathcal{O}(s) \right]
= -\frac 2{\pi}  \left[ \frac 1{2s}+
\ln \left(\frac\mu\Omega\right) + \mathcal{O}(s) \right] \label{Es0}.
\end{equation}
The pole term
\begin{equation}
E_{\rm pole}=-\frac 1{\pi s} \,\frac {m}{\sqrt 2} \label{Epole}
\end{equation}
(where we restored the $m$-dependence by using dimensional
considerations)  can be removed by an infinite renormalization of
the mass
\begin{equation}
\delta m^2=\frac \lambda{2s\pi} \,,\label{delm2}
\end{equation}
c.f. (\ref{clen})\footnote{In principle one can do renormalization
directly in the action by using (\ref{a2ki}). The heat kernel
expansion for NC $\phi^4$ constructed in \cite{Vassilevich:2005vk}
predicts the same multiplier of $2/3$ which relates counterterms
in commutative and non-commutative cases. On should however keep in 
mind that the results of \cite{Vassilevich:2005vk} are valid for
background fields which decay rapidly at the infinity. This is not
the case of the kink solution, which tends to {\it different}
constants at two infinities. Sensitivity of the heat kernel
expansion to the asymptotic behavior of background fields
is a generic feature of NC manifolds. The only exception are
the expansions for operators containing only left or only
right Moyal multiplications. In this latter case universal formulae
exist as long as corresponding integrals are convergent
\cite{Vassilevich:2003yz,Gayral:2004ww}.}. 
This counterterm does not depend on $\theta$ but
is $2/3$ of the corresponding counterterm in the commutative case
\cite{Bordag:1994jz,Bordag:2002dg}.

After removing the pole one can lift the regularization taking the
limit $s\to 0$.
\begin{eqnarray}
&&E=\frac 12 \sum \omega_B + \int_M^\Omega d\omega\, \omega \rho
(\omega) \nonumber\\
&&\qquad +\int_\Omega^\infty d\omega\, \omega (\rho
(\omega)-\rho_{\rm as}(\omega)) -\frac 2\pi \ln \left( \frac
{\mu}{\Omega} \right)\label{Efin}
\end{eqnarray}
We remind that $\rho(\omega)=(2\pi)^{-1}\partial_\omega \delta
(\omega)$. 

The presence of a free parameter $\mu$ reflects the
possibility of a finite renormalization.  To fix $\mu$ one needs
a normalization condition. In the commutative case in $(1+1)$ dimensions
it is usually required that the tadpoles are cancelled by counterterms
(the ``no-tadpole'' condition). This condition is formulated on
a constant topologically trivial background. In noncommutative theories
one-loop divergences on a constant background coincide with the
divergences in the commutative case and thus differ from the divergences
in the kink sector. (The difference is precisely the $2/3$ factor
discussed above). Therefore, the no-tadpole condition is not suitable
for us. In the case of commutative kink, there is the large mass subtraction
scheme \cite{Bordag:1998vs} which is equivalent to the no-tadpole condition
\cite{Bordag:2002dg}.
Although this scheme can be applied even in non-renormalizable theories,
it is not clear how to implement it in the noncommutative case.
This situation is not hopeless, but it is more natural to address it
together with studying the commutative limit $\theta\to 0$. Our
numerical scheme (see below) does not work well in this limit. Therefore,
we postpone the discussion of physically motivated normalization
conditions until a future publication. Here, for the sake of
simplicity, we put
\begin{equation}
\mu=M. \label{muM}
\end{equation}
The values of $E$ for other choices of $\mu$ differ by a shift
$-(2/\pi)\ln (\mu/M)$.
\section{Calculation of energy}\label{sec-cal}
\begin{figure}\label{fig3}
\includegraphics[width=\linewidth]{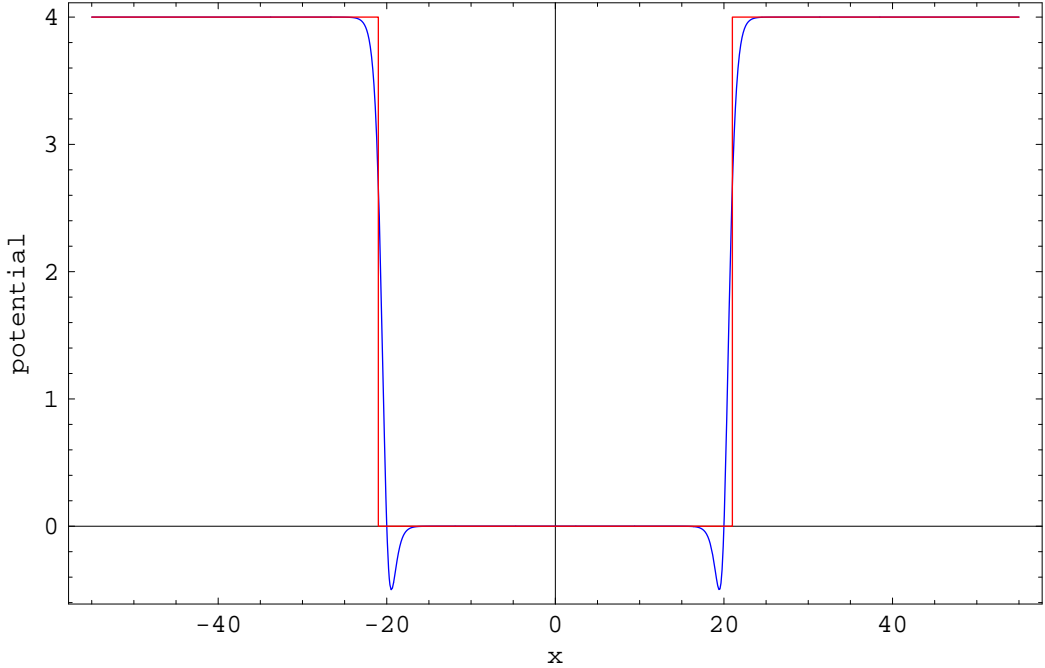}
\caption{The effective potential $U(x, \omega)+M^2$ for $\omega=2$, 
$\theta=10$ and its square well approximation.}
\end{figure}

Let us start from the first term in the formula (\ref{Efin}) which contains the summation over the bound state frequencies.
Here we shall distinguish the two cases: first, for $ \theta \omega $ considerably larger than $1$, and second, 
$ \theta \omega $ comparable or smaller than $1$.
When $ \theta \omega \gg 1$ we have very good approximation by a square well potential (cf. Fig.\ \ref{fig3}).
The square well potential is, $V=0$ for $|x|>l$ and $V=-V_0<0$ for
$|x|<l$. (In our case $V_{0}= 4$). The scattering data for this
potential read
\begin{eqnarray}
&&s_{21}=s_{12}=\frac{V_0 e^{-2ikl} (e^{2i \omega  l}-e^{-2i \omega  l})}
{(k+ \omega )^2e^{-2i\omega l}-(k-\omega)^2e^{2i \omega l}} \\
&&s_{11}=s_{22}=\frac{4\omega k e^{-2ikl}}
{(k+ \omega )^2e^{-2i \omega l}-(k-\omega )^2e^{2i \omega  l}}
\end{eqnarray}
with $   \omega = \sqrt{k^2+V_0}$. The width of an approximating square-well 
potential 
must be chosen such that it correctly reproduces the leading asymptotics
of the phase shift, i.e. through the condition
\begin{equation*}
-2l(\omega) V_0=\int dx U(x;\omega) = -4(2 + 2 \theta \omega 
{\rm Coth}(2 \theta \omega)).
\end{equation*}
As $2l(\omega) = \theta \omega + c $, it immediately follows $c=1$,
cf. Fig.\ \ref{fig3}.
(Note, that in \cite{Vassilevich:2003he} the value $c=0$ was taken
to estimate the number of bound states for large $\theta$. To the leading
order in $\theta$ both approximations coincide, but the one chosen here
reproduces the scattering data of $U(x;\omega)$ with a better accuracy.)

It is well known from quantum mechanics that the bound states of the 
square well potential obey the equations 
\begin{equation}
\tan(\theta \omega^2) -\frac{\sqrt{4 - \omega^2}}{\omega}=0, \quad \tan^{-1}(\theta \omega^2)+\frac{\sqrt{4 - \omega^2}}{\omega}=0.\label{Bound States}
\end{equation} 
In the Table I. one can see some bound states calculated with the above formulas (\ref{Bound States}) for $\theta=10$. 

Alternatively, for finding the bound states one can use the 
well-known WKB formula of the first order
\begin{equation}
\int_{x_{1}}^{x_{2}} \sqrt{-U(x, \omega_{n})+\omega_{n}^2} d x = \pi (n+ \frac{1}{2}),\label{Bound States 2}
\end{equation} 
where $x_{1}$ and $x_{2}$ are the turning points. It is well-known that WKB approach works well for the low laying bound states, i.e. the smaller $\theta \omega$, 
the better accuracy of the WKB approximation. At the same time, the larger  $\theta \omega$, the better accuracy of the square well approximation. As a result, 
one can see in the Table I, with an example of $\theta =10$ case, that highly excited bound state frequencies obtained from the square well approximation agree 
very well with their WKB values in the regime of large $\theta \omega$. Thus the difference between the WKB and SW (square well) data is less than fractions 
of one percent. Therefore we expect that relative error of our calculations of energy for large $\theta$ should not exceed one percent.
Let us also remind  that the zero mode $\omega_{0}=0$ is the same as 
in the commutative case. (This is the translation zero mode).
To summarize, for all bound state frequencies $\omega_{n\ge 1}$ we use the
WKB approximation, which is known to be accurate for lower eigenstates, and
which practically coincides with the frequencies obtained from the square well
approximation near the barrier.

On the contrary to the discrete spectrum, when considering integrals over continuous part of the spectrum (second and third terms  in (\ref{Efin})),
we start from $\theta =2$, $\omega =2$ and can use the  square well approximation. In other words $\Omega=M$ in (\ref{Efin}), if we are limited by not small 
values of $\theta$. For small $\theta$  and small $\omega$ (of the bound states and in the beginning of the continuous part of the spectrum) 
the  the potential does not look like square well, but rather like a modified 
P\"oschl-Teller potential. The WKB method is certainly justified for that case as well for calculation of bound states, 
yet is computationally difficult for the continuous spectrum. That is why we did not consider the limit of small $\theta$.

We used the following expression for the phase shift
\begin{equation}
\delta = \frac{1}{2 i} \ln(s_{11}^2-s_{21}^2)
\end{equation} 
(cf. Appendix \ref{app-ve}).
The dominant asymptotic behavior of $\rho(\omega)_{as}$ at large $\omega$ 
is given by (\ref{rhoas}) above.

Since we put $\Omega=M$ the second term in (\ref{Efin}) vanishes.
The calculation of the third term in (\ref{Efin}) was done by numerical integration with a $\rho$ function given by a square well
approximation. The numerical integration was performed by {\it Mathematica} and the values for different $\theta$ are given in Fig. 1.
One can see there that the roughly linear dependence on $\theta$ takes place.
The approximating square well potential differs from the exact potential
$U(x;\omega)$ only slightly near the points $x=\pm (\theta\omega+1)$,
and the form of this difference practically does not depend on $\theta$.
Therefore, especially since we deal with a massive field, we may hope
that the total error will remain small and bounded independently of
$\theta$, so that our conclusion about the linear growth of the
integral over the continuous spectrum will remain true even if a better
approximation is used. Besides, as we shall see, the contribution of the
continuous spectrum is about $1/2$ of the contribution of the bound
states, so that any error in the continuous spectrum is less important.
Adding the values for the integration over continuous spectrum (Fig. 1) to the sum over the bound state frequencies (see Table I, except for the
case $\theta=7$ which is not presented explicitly in order not too overload
the Table), we get
\begin{eqnarray}
&&E = 2.76, \quad (\theta =2)\nonumber\\
&&E = 2.95, \quad (\theta =3)\nonumber\\
&&E = 4.66, \quad (\theta = 5)\label{Etheta}\\
&&E = 6.49, \quad (\theta = 7)\nonumber\\
&&E = 9.16, \quad (\theta = 10)\nonumber
\end{eqnarray}
Let us note that as the contribution to the energy from the integral over continuous spectrum is linear with $\theta$ (see Fig. 1) and the contribution to 
the bound state is linear as well, the final values for $E$ as a function of $\theta$ is linear in $\theta$ (see Fig. 2).

In the previous section we put $\mu=M$. Values of the vacuum energy for
a different choice of $\mu$ are obtained from the one given in (\ref{Etheta})
by a constant shift, $E\to E-(2/\pi)\ln (\mu/M)$. We fixed the mass units
so that $m=\sqrt{2}$. In arbitrary units $E$ should be multiplied by 
$m/\sqrt{2}$.

\begin{figure}\label{fig1}
\includegraphics[width=\linewidth]{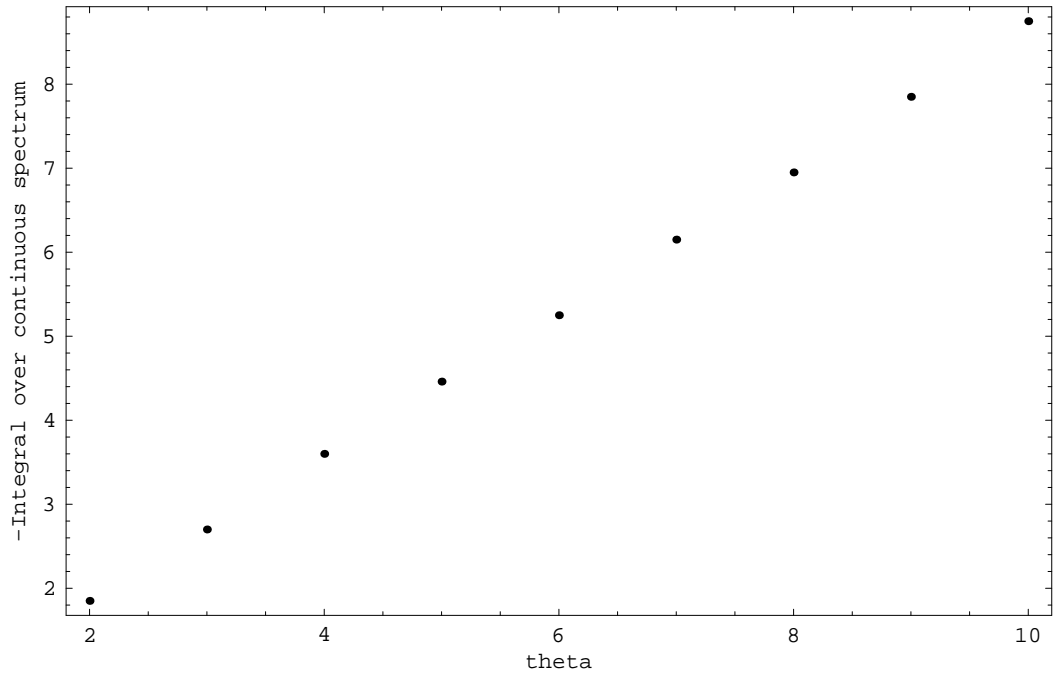}
\caption{The values of the integral $-\int_M^\infty d\omega\, \omega (\rho
(\omega)-\rho_{\rm as}(\omega))$ as a function of $\theta$.}
\end{figure}

\begin{figure}\label{fig2}
\includegraphics[width=\linewidth]{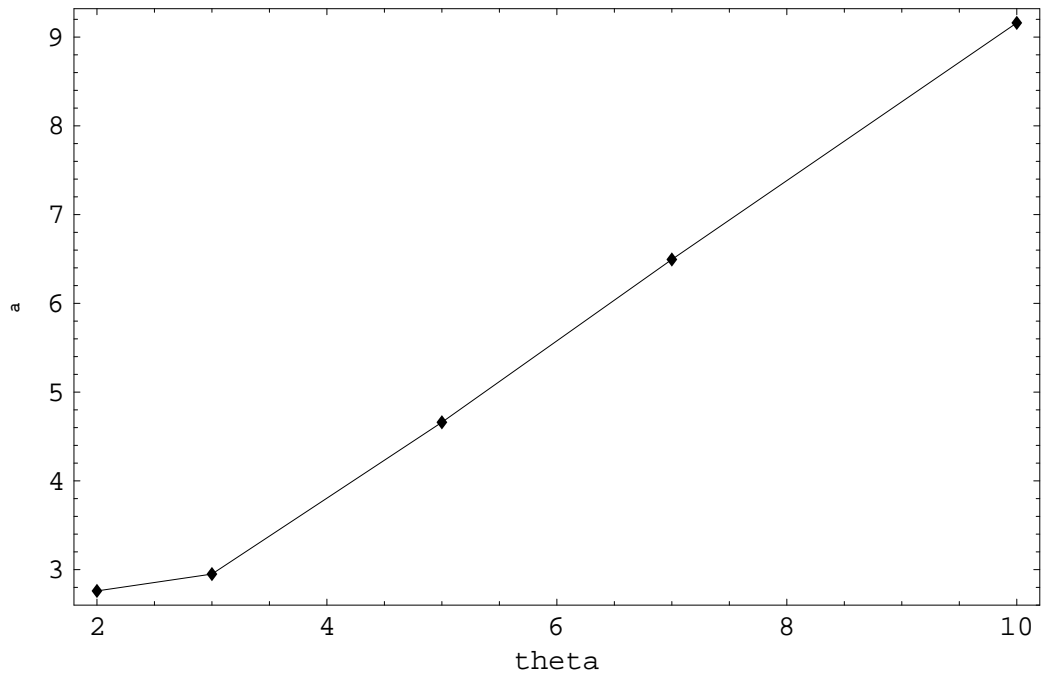}
\caption{The values of the $E$ as a function of $\theta$. The angle of the slope of the line is: $\arctan \alpha = 0.726$.}
\end{figure}

\begin{table}
\begin{center}
\begin{tabular}{ccccc}
\hline
$\theta =2 (WKB)$ & $\theta=3 (WKB)$ &  $\theta=5 (WKB)$        &  $\theta=10 (WKB)$       \\
\hline
\hline
$\omega_{1}=0.955$  & $\omega_{1}=0.790$  &$\omega_{1}=0.613 $ &  $\omega_{1} = 0.433   $  \\
\hline
$\omega_{2}=1.263$  & $\omega_{2}=1.056$  &$\omega_{2}=0.826$  &  $\omega_{2} = 0.586   $  \\
\hline
$\omega_{3}=1.496$  & $\omega_{3}=1.260$  &$\omega_{3}=0.995$ &  $\omega_{3}=  0.708  $  \\
\hline
$\omega_{4}=1.687$  & $\omega_{4}=1.143$  &$\omega_{4}=1.135$  &  $\omega_{4} = 0.811  $   \\
\hline
$\omega_{5}=1.847$  & $\omega_{5}=1.579$ &$\omega_{5}=1.258$ &  $\omega_{5} = 0.901  $   \\
\hline
$\omega_{6}=1.970$  & $\omega_{6}=1.712$  &$\omega_{6}=1.368$ &   $\omega_{6} =0.983  $   \\
\hline
--                     & $\omega_{7}=1.828$  &$\omega_{7}=1.469$ &  $\omega_{7}=  1.058  $   \\
\hline
--                     & $\omega_{8}=1.930$  &$\omega_{8}=1.564$ &  $\omega_{8} = 1.128 $    \\
\hline
--                     & --&$\omega_{9}=1.650$ &   $\omega_{9} =1.194   $  \\
\hline
--                     & --&$\omega_{10}=1.737$ &    $\omega_{10}= 1.256  $   \\
\hline
--                     &-- &$\omega_{11}=1.810$ &  $\omega_{11} = 1.314  $  \\
\hline
--                     & --&$\omega_{12}=1.880$ &  $\omega_{12} = 1.371 $     \\
\hline
--                     &-- &$\omega_{13}=1.946$ &      $\omega_{13}= 1.425  $     \\
\hline
--                     &-- &--    &   $\omega_{14} =1.476  $     \\
\hline
--                     &-- &--     &  $\omega_{15} = 1.526  $     \\
\hline
--                     &-- &--     &  $\omega_{16} = 1.574  $    \\
\hline
--                     &-- &--      & $\omega_{17}  = 1.621  $   \\
\hline
--                     &-- &--      &  $\omega_{18} = 1.665  $   \\
\hline
--                     &-- &--     &  $\omega_{19} = 1.709  $    \\
\hline
--                     &-- &--     &  $\omega_{20} = 1.751  $   \\
\hline
--                     &-- &--     &  $\omega_{21} = 1.793  $  \\
\hline
--                     &-- &--     &  $\omega_{22} = 1.833  $  \\
--                     &-- &--     &  $\omega_{22SW} = 1.828 $  \\
\hline
-                     &-- &--     &  $\omega_{23} = 1.872 $    \\
--                     &-- &--     &  $\omega_{23SW} = 1.869 $ \\
\hline
--                     &-- &--     &  $\omega_{24} = 1.908  $     \\
--                     &-- &--     &  $\omega_{24SW} = 1.909 $ \\
\hline
--                     &-- &--     &  $\omega_{25} = 1.944  $  \\
--                     &-- &--     &  $\omega_{25SW}= 1.948 $ \\
\hline
--                     &-- &--     &  $\omega_{26} = 1.978$   \\
--                     &-- &--     &  $\omega_{26SW} = 1.985$ \\
\hline
\hline
\end{tabular}
\caption[smallcaption]{Values of the bound state frequencies computed by formula (\ref{Bound States}) (only for the highest five states for $\theta=10$) 
and WKB frequencies found by (\ref{Bound States 2}). The comparison shows that 
the WKB approximation results differ from the SW (square well) approximation 
only by fractions of one percent. }
\label{label}
\end{center}
\end{table}

\newpage
\section{Conclusions}
In this work we found that the divergences in the zeta-regularized
one-loop vacuum energy of NC kink  (defined as a half-sum over 
egenfrequencies) can be removed by the mass renormalization. 
This renormalization is, however, different from the one
required in the topologically trivial sector.
Although the effective potential which defines the spectrum
of excitations above the kink depends on frequencies, for
intermediate and large values of the NC parameter $\theta$ the finite
part can be calculated by a combination of the WKB method and
an approximation by a square well potential. For large noncommutativity
the one-loop vacuum energy grows linearly with $\theta$, so that sooner or
later it should become larger than the classical value thus
signalling break-down of the perturbative expansion.

Our results may be improved and extended in a number of ways. First of
all, we need a method of calculations which would work for small
$\theta$ and a physically motivated normalization condition to
fix the parameter $\mu$. Having these ingredients at hand, one can 
address the question whether quantum corrections to the NC kink
are smooth in the limit $\theta\to 0$ and whether they reproduce
the commutative result \cite{Dashen:1974cj} in this limit.
\appendix
\section{Vacuum energy in the zeta regularization}\label{app-ve}
Here we derive eq.\ (\ref{EC2}) by using the approach of Bordag
\cite{Bordag:1994jz} and making necessary modifications due to
the noncommutativity. In our exposition we also use Ref.\ 
\cite{Bordag:2002dg}. First we introduce a cut-off at large distances
by imposing the Dirichlet conditions on the perturbations 
$\eta (-L(k))=\eta (L(k))=0$. In commutative case, when
the potential has no dependence on $\omega$, it is enough to
take $L(k)=const.$ In our case the potential does depend on $\omega$.
We would like to ensure that $x_\pm$ are far away from the boundary
for all $\omega$. This can be achieved by taking $L(k)=\theta\omega
+L_0$ where $L_0$ is a large positive constant. Later we shall
consider the limit $L_0\to\infty$. The frequency dependent boundary
condition is the main novelty here as compared to previous works.
We shall demonstrate that it does not change the result.

Without boundaries for each momentum
$k$ there are two independent solutions $\eta_1$, $\eta_2$ 
of the wave equation with the asymptotic
behavior
\begin{eqnarray}
&&\eta_1\sim e^{ikx} + s_{12} e^{-ikx} ,\quad \eta_2\sim s_{22}e^{ikx}
\quad \mbox{for} \ {x\to -\infty}\nonumber\\
&&\eta_1\sim s_{11} e^{ikx} ,\quad \eta_2\sim s_{21}e^{ikx}+
e^{-ikx}
\quad \mbox{for} \ {x\to \infty}.\label{asbeh}
\end{eqnarray}
The potential $U$ is symmetric under the reflection $x\to -x$.
Consequently $s_{11}=s_{22}$, $s_{21}=s_{12}$.
For large but finite
$L_0$ the spectrum is discrete and is defined by the condition
\begin{equation}
f(k)=((s_{11}+s_{21})e^{ikL}+e^{-ikL})
((s_{11}-s_{21})e^{ikL}-e^{-ikL})
=0,\label{fk}
\end{equation}
where the bracket with the plus (resp., minus) sign corresponds to
a symmetric (resp., antisymmetric) solution.

It is known that if we have discrete spectrum only, the zeta-regularized
vacuum energy is a sum over the eigenfrequencies, $E=\frac 12 \sum_n
\omega_n^{1 -2s}=\frac 12 \sum_n (k_n^2+M^2)^{\frac 12 -s}$. The function
$\partial_k \ln f(k)$ has poles at $k=k_n$ with unit residues.
Therefore, we can rewrite regularized $E_C$ in the form of a contour
integral,
\begin{equation}
E_C^{(L_0)}=\frac 12 \oint \frac{dk}{2\pi i} (k^2+M^2)^{\frac 12 -s} 
\frac \partial{\partial k} \ln f(k).\label{AEC1}
\end{equation}
The integration contour runs anticlockwise around the real positive
semiaxis and consists of one branch at $k={\rm Re}\, k+i\epsilon$,
a second branch at $k={\rm Re}\, k-i\epsilon$, and a small segment
$-\epsilon \le {\rm Im}\, k \le \epsilon$ along the imaginary axis.
Along the upper part of the contour we keep in $f(k)$ only the terms
with $\exp (-ikL(k))$
since $\exp (ikL(k))$ vanishes as
$L_0\to \infty$. Along the lower part of the contour we retain
$\exp (ikL(k))$. The contribution from the
third part can be dropped, as in \cite{Bordag:1994jz,Bordag:2002dg}. We have
\begin{equation}
E_C^{(L_0)}=\frac 12 \int_0^\infty 
\frac{dk}{2\pi i} (k^2+M^2)^{\frac 12 -s} 
 \frac \partial{\partial k} 
(4ikL(k) +\ln (s_{11}^2-s_{22}^2)).\label{AEC2}
\end{equation}
Next we take into account 
\begin{equation}
s_{11}^2-s_{21}^2=e^{2i\delta(k)},\label{ssdel}
\end{equation}
where $\delta (k)$ is the phase shift, and
subtract the contribution from free fields of mass $M$
satisfying the same boundary conditions (i.e., the expression
(\ref{AEC2}) with $\delta (k)=0$). After taking the limit $L_0\to\infty$
 we obtain eq.\ (\ref{EC2}) for the vacuum energy.

Note, that taking the boundaries into account explicitly is essential
in supersymmetric theories (where the boundary conditions must be 
supersymmetric) \cite{Bordag:2002dg}. 

\section*{Acknowledgments}
R. K. was supported by {\it Japan Society for Promotion of Science (JSPS)}, 
Japan. D.V.V.\ thanks FAPESP for the support.

\end{document}